\documentclass[10pt,a4paper,5p, oneside, preprint]{elsarticle}

\pdfoutput=1

\usepackage[english]{babel}
\usepackage{blindtext}
\usepackage[pdftex,plainpages=false,pdfpagelabels]{hyperref}
\usepackage{amsmath,amssymb,amsthm,amstext,amsfonts}
\usepackage{color}
\usepackage{units}
\usepackage[latin1]{inputenc}
\usepackage{acronym}
\usepackage{subfigure}
\usepackage{booktabs,longtable}
\usepackage{enumitem}
\usepackage[switch]{lineno}

\newcommand*\patchAmsMathEnvironmentForLineno[1]{%
  \expandafter\let\csname old#1\expandafter\endcsname\csname #1\endcsname
  \expandafter\let\csname oldend#1\expandafter\endcsname\csname end#1\endcsname
  \renewenvironment{#1}%
     {\linenomath\csname old#1\endcsname}%
     {\csname oldend#1\endcsname\endlinenomath}}%
\newcommand*\patchBothAmsMathEnvironmentsForLineno[1]{%
  \patchAmsMathEnvironmentForLineno{#1}%
  \patchAmsMathEnvironmentForLineno{#1*}}%
\AtBeginDocument{%
\patchBothAmsMathEnvironmentsForLineno{equation}%
}

\newacro{UHE}{ultra-high-energy}
\newacro{GZK}{Greisen-Zatsepin-Kuzmin}
\newacro{CMB}{cosmic microwave background}
\newacro{UHECR}{UHE cosmic rays}
\newacro{LPM}{Landau-Po\-me\-ran\-chuk-Migdal}
\newacro{LDF}{lateral distribution function}
\newacro{MC}{Monte Carlo}
\newacro{MVA}{Multi-Variate Analysis}
\newacro{CIC}{Constant Intensity Cut}

\graphicspath{{figures/}}

\begin{document}

\title{$F_\gamma$: a new observable for photon-hadron discrimination
  in hybrid air shower
events}
\author[si]{M. Niechciol\corref{cor1}}
\ead{niechciol@physik.uni-siegen.de}
\author[si]{M. Risse}
\author[si]{P. Ruehl}
\author[si,lpnhe]{M. Settimo\fnref{fn1}}
\author[si]{P. W. Younk\fnref{fn2}}
\author[si,iteda]{A. Yushkov\fnref{fn3}}

\address[si]{Universit\"at Siegen, Department
  Physik, Siegen, Germany}
\address[lpnhe]{Laboratoire de Physique Nucl\'eaire et de Hautes
  Energies (LPNHE), Universit\'es Paris 6 et Paris 7, CNRS-IN2P3,
  Paris, France}
\address[iteda]{Instituto de Tecnolog\'ias en Detecci\'on y
  Astropart\'iculas (CNEA, CONICET, UNSAM), Buenos Aires, Argentina}

\cortext[cor1]{Corresponding author}
\fntext[fn1]{Now at SUBATECH, CNRS/IN2P3, Universit\'e de Nantes,
  \'Ecole des Mines de Nantes, Nantes, France}
\fntext[fn2]{Now at Physics Division, Los Alamos National Laboratory, Los Alamos, NM, USA}
\fntext[fn3]{Now at Institute of Physics (FZU) of the Academy of Sciences of the Czech Republic, Prague, Czech Republic}

\begin{abstract}
To search for ultra-high-energy photons in primary cosmic rays, air shower observables are needed that
allow a good separation between primary photons and primary
hadrons. We
present a new observable, $F_\gamma$, which can be extracted from ground-array
data in hybrid events, where simultaneous measurements of the longitudinal
and the lateral shower profile are performed. The observable is based
on a template fit to the lateral distribution measured by the ground
array with the template taking into account the complementary information from the
measurement of the longitudinal profile, i.e. the primary energy and
the geometry of the shower. $F_\gamma$ shows a very good photon-hadron
separation, which is even superior to the  separation given
by the well-known $X_\text{max}$ observable
(the atmospheric depth of the shower maximum). At energies around
$\unit[1]{EeV}$ ($\unit[10]{EeV}$), $F_\gamma$ provides a background
rejection better than $\unit[97.8]{\%}$ ($\unit[99.9]{\%}$) at a
signal efficiency of $\unit[50]{\%}$. 
Advantages of the observable $F_\gamma$ are its technical stability
with respect to irregularities in the ground array (i.e. missing or
temporarily non-operating stations) and that it can be applied over the full
energy range accessible to the air shower detector, down to its threshold energy.
Furthermore, $F_\gamma$ complements nicely to $X_\text{max}$ such that
both observables can well be combined to achieve an even better
discrimination power, exploiting the rich information available in hybrid events.
\end{abstract}


\maketitle


\section{Introduction}
\label{sec:introduction}

The discovery of \ac{UHE} photons, i.e. photons with an
energy larger than ${\sim}\unit[10^{18}]{eV} = \unit[1]{EeV}$, in primary cosmic rays would be of
particular interest not only for the field of astroparticle physics,
but also for related fields such as particle physics, astrophysics and fundamental physics~\cite{risse07}. For example, UHE
photons are tracers of the \ac{GZK} process, i.e. the interactions of
\ac{UHE} protons, propagating through the Universe, with photons from
the \ac{CMB}. In these interactions, neutral pions are produced via
the Delta resonance ($p + \gamma_\text{CMB} \rightarrow \Delta^+ \rightarrow p + \pi^0$).
These pions subsequently decay into pairs of \ac{UHE} photons with energies
at typically $\unit[10]{\%}$ of the energy of the primary \ac{UHE}
proton. If these predicted \ac{GZK} photons are observed on Earth, it
would be an indicator for the \ac{GZK} process being the reason for
the observed suppression in the energy spectrum of
\ac{UHECR}~\cite{auger08a}. Observing \ac{UHE} photons could also help
to pinpoint the very sources of \ac{UHECR}, since photons, unlike 
charged cosmic rays, are not deflected by magnetic fields. The attenuation length for photons in the EeV range varies between some
$\unit[100]{kpc}$ at $\unit[1]{EeV}$ and a few Mpc at
$\unit[10]{EeV}$~\cite{auger17b}, encompassing possible galactic and
nearby extragalactic sources. The detection of \ac{UHE} photons is of great interest for
fundamental physics as well. For instance, the registration of a single photon
in the EeV range could improve existing bounds on Lorentz invariance
violation in the context of a modified Maxwell theory by
several orders of magnitude~\cite{klinkhamer10}. In addition,
observing the particle cascade initiated by a \ac{UHE} photon in the
atmosphere allows testing particle interactions at extreme
energies and searching for new physics~\cite{risse07,diaz16}.\\

\begin{figure}[t!]
	\centering
		\includegraphics[width=\columnwidth]{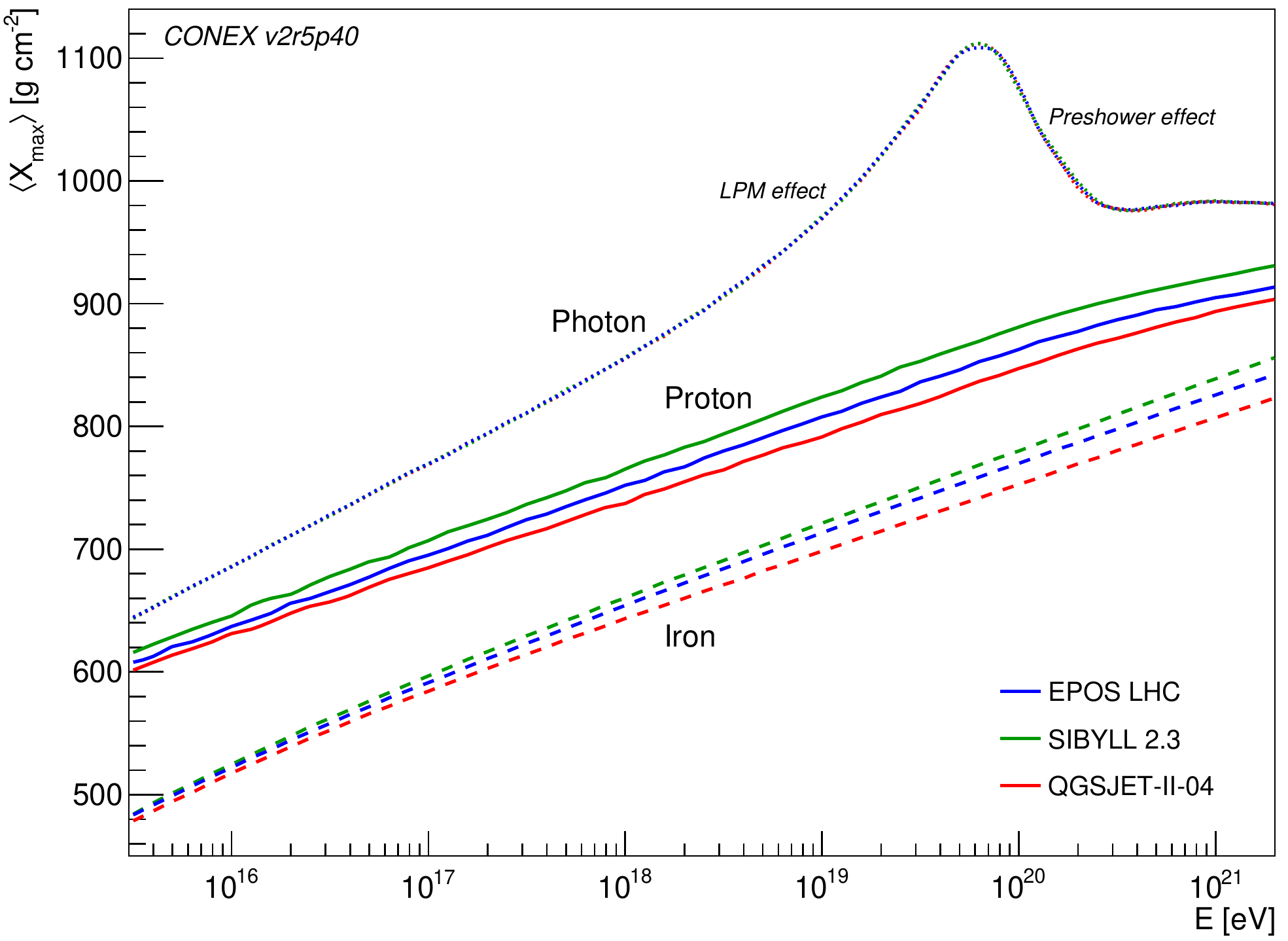}
	\caption{Average atmospheric depth of the shower maximum,
          $\left<X_\text{max}\right>$, as a function of the primary
          energy for simulated air
          showers initiated by photons, protons and iron nuclei. The simulations have been done using the
          Conex simulation code~\cite{bergmann06,pierog04} with three different
          hadronic interaction models (EPOS~LHC~\cite{pierog15}, SIBYLL~2.3~\cite{ahn09,riehn15} and QGSJET-II-04~\cite{ostapchenko11}). For the
          calculation of the preshower effect, the location of the
          Pierre Auger Observatory in Argentina has been used.}
	\label{fig:xmaxplot}
\end{figure}

Due to their small incoming flux (less than one particle per
square kilometer per year), \ac{UHE} cosmic particles impinging on the
Earth can only be detected indirectly through the measurement of the
air showers they initiate when entering the Earth's
atmosphere. For the identification of primary photons in the recorded
air shower data, the challenge is to separate photon-induced showers from those
initiated by hadrons.
Thus, the differences
between these two classes of air showers are of great importance. Air showers initiated by
\ac{UHE} photons develop, on average, deeper in the atmosphere than air
showers of the same primary energy initiated by hadrons. This can be
expressed through the observable $X_\text{max}$, which describes the
atmospheric depth of the shower maximum (see Fig.~\ref{fig:xmaxplot}). At
larger energies, additional effects like the \ac{LPM} effect and preshower
processes above the atmosphere, which further influence the shower
development, have to be taken into account~\cite{risse07}. $X_\text{max}$ has become a key observable for
current cosmic-ray research, mostly due to the fact that it
can be accessed directly using the air-fluorescence technique. Current
air shower experiments like the Pierre Auger Observatory~\cite{auger15} or the
Telescope Array~\cite{ta12} are following a hybrid approach, where
fluorescence detectors are complemented by a ground array of particle
detectors. The typical $X_\text{max}$ resolution of e.g. the Pierre
Auger Observatory is better than $\unit[26]{g\,cm^{-2}}$ at energies
around $\unit[10^{17.8}]{eV}$, improving with energy to about
$\unit[15]{g\,cm^{-2}}$ above $\unit[10^{19.3}]{eV}$~\cite{auger14b}.
This is much smaller than the differences in the average $X_\text{max}$
between photon- and hadron-induced air showers at these energies
($\sim\!\unit[100]{g\,cm^{-2}}$ at $\unit[10^{18}]{eV}$ and
$\sim\!\unit[160]{g\,cm^{-2}}$ at $\unit[10^{19}]{eV}$, cf. Fig.~\ref{fig:xmaxplot}).\\

To fully exploit this hybrid approach
and to improve the photon-hadron separation power,
it is useful to
complement the observable $X_\text{max}$ with an additional observable
that is based on the data from the ground array. In the past, observables such as the curvature of the shower front or the risetime
of the signals in the detectors of the ground array have been used in
the context of the search for photons~\cite{auger08b,ta13}. However, such
observables can only be reliably estimated when a minimum number of detectors
of the ground array have triggered. For example, at least five detectors
are needed to determine the curvature of the shower front with
acceptable accuracy. This effectively places a lower limit---for past analyses typically at $\unit[10]{EeV}$---on the energy region that can be accessed with an analysis based
on these observables. At lower energies, i.e. in the EeV range, the
observable $S_b$~\cite{ros11} has been successfully used in past
analyses~\cite{settimo11,auger14,auger17a,auger17b}. $S_b$ is based on
the sum of the signals $S_i$, measured in the individual
detectors $i$ of the ground array, weighted by the distances $r_i$ of the
detectors to the shower axis:
\begin{linenomath}
\begin{equation}
S_b = \sum_i S_i \left(\frac{r_i}{\unit[1000]{m}}\right)^b,
\end{equation}
\end{linenomath}
with the free parameter $b$, which has to be fixed for a given analysis ~\cite{ros13}.

 Air showers initiated by photons have, on average, a smaller $S_b$ than air showers induced by
hadrons of the same primary energy~\cite{ros13}. However, an observable like $S_b$ can be
significantly affected by any incompleteness in the detector
array, which would lead to an underestimation of the true value.
Such an incompleteness may be due to the borders of the array, due to missing
detectors (important e.g.\ during the deployment phase of the array), or
due to temporarily non-operating detectors.
As an example for the latter effect, let us assume that at any time, $\unit[1]{\%}$ of
the detectors from the ground array are temporarily
non-operating. For an array geometry where
the detectors are arranged on a hexagonal grid, this means that
$\unit[6]{\%}$ of all events contain at least one non-operating detector in the
first hexagon around the detector measuring the largest signal. When
also the second hexagon is considered, about $\unit[18]{\%}$ of all measured
events are affected. Special care must then be taken, e.g. in the event
selection, to prevent an underestimation of the $S_b$ value
for a given event due to these incompletenesses, which could mimic the expected behaviour of a
photon-induced air shower. This holds especially at energies not far
from the energy threshold of the experiment, where usually only very few detectors of the ground array are
triggered and an omission of a signal from a detector can alter 
the $S_b$ value substantially.\\

In this paper, we
describe a new observable, called $F_\gamma$,
which can be used at all energies down to the threshold set by the shower array.
This observable exploits,
similarly to $S_b$, the lateral distribution of the density of secondary
particles from the air shower on ground level and is, as will be
discussed later, complementary to $X_\text{max}$.
Unlike $S_b$, missing stations will not alter the central value of $F_\gamma$
(but only increase its uncertainty), leading to an improved stability of
the observable.
 The lateral
distribution can be described by a \ac{LDF}, which can be determined
from the ground-array data. The shape of the \ac{LDF} depends on the
type of the primary particle initiating the air shower: for
photon-induced air showers, which on average exhibit a smaller number of
secondary muons and a larger $X_\text{max}$ compared to hadron-induced air showers of the same
primary energy, the \ac{LDF}
is steeper, leading to a smaller signal in the detectors of the ground
array compared to the signal measured from hadron-induced air showers
at the same distance from the shower core. $F_\gamma$ is based
on a fit of the \ac{LDF}. Therefore, it is largely unaffected by incompletenesses
of the ground array. In addition, $F_\gamma$ can be determined for
events with very few triggered detectors in the ground array (even
just a single one), hence it
can be applied also to lower-energy events, where observables such as
the curvature of the shower front cannot be determined.\\

In the following sections, we
describe how the observable $F_\gamma$ is defined and evaluate the
performance of the observable in distinguishing photon-induced air
showers from those induced by hadrons by using \ac{MC} simulations of
air showers.


\section{Description of the observable}
\label{sec:description}

\subsection{General idea}

The observable $F_\gamma$ is based on a template fit of an \ac{LDF}
to the data recorded by the ground array. In this fit, we assume the
primary particle initiating the air shower was a photon, and we
determine the expected signal recorded in a detector of the ground
array at a reference distance under
this assumption. In this fit, two things will be different between
photon- and hadron-induced showers with the same primary
energy. First, the LDF fit will better describe the lateral particle
distributions for primary photons than for primary hadrons. Second, the 
signal at a certain distance from the shower core will be smaller for showers initiated by photons than for those
initiated by hadrons, as is known from air shower physics. We
exploit the second difference. We normalize the expected signal
obtained from the fit to the average signal expected for
hadron-induced air showers
of the same primary energy and zenith angle, and we call the resulting
quantity $F_\gamma$. In short, the sequence to determine $F_\gamma$ for
a given air shower event is as follows:
\begin{itemize}[noitemsep,nolistsep]
\item[-] From hybrid observations, in particular of the longitudinal shower profile,
the energy $E$ and zenith angle $\theta$ of the event are reconstructed;
\item[-] Using $E$ and $\theta$ as an input, a photon LDF template is fit to the
ground-array data of the event. The template is prepared in advance
using extensive photon simulations.
The fit determines the LDF normalization $S_{1000|\gamma}$,
the only remaining free parameter of the template;
\item[-] The average ground signal
  $\left<S_{1000}\right>$ in case of hadron-induced air showers expected for this event is calculated using $E$ and $\theta$. The relation between these
  quantities is known from the standard energy calibration of the ground array;
\item[-] $F_\gamma$ for this event is then given by the ratio $\frac{S_{1000|\gamma}}{\left<S_{1000}\right>}$.
\end{itemize}

\subsection{Specific implementation}

In the following section, we describe a specific realization of the
observable $F_\gamma$. In particular, we use a given functional
form to describe the \ac{LDF}. However, the core concept of the
observable as described above does not depend on the very choice of
a functional form for the
\ac{LDF} or even a particular experimental setup, but it can be applied to
different functional forms and adapted to different experiments.\\

To describe the \ac{LDF}---i.e. the signal $S$ measured in a detector
at a perpendicular distance $r$ from the shower axis---, we use an NKG-type function, which has the
following form:
\begin{equation}
S(r) = k \left(\frac{r}{r_\text{opt}}\right)^{\beta} \left(\frac{r + r_\text{s}}{r_\text{opt}+r_\text{s}}\right)^{\beta},
\end{equation}
where $r_\text{opt}$ is the optimum distance at which the characteristic
parameters of the air shower can be determined to reduce
e.g. uncertainties due to a lack of knowledge about the true \ac{LDF},
$r_\text{s}$ is a scaling parameter, $k$ is a normalization parameter
equal to the signal at the optimum distance, and $\beta$ is
the slope of the \ac{LDF}. The choice of the optimum distance depends
on the geometrical properties of the ground array. For the
experimental setup of
the Pierre Auger Observatory---which is assumed from now on---with its hexagonal grid and a spacing
of $\unit[1500]{m}$ between the individual detectors, $r_\text{opt} \simeq
\unit[1000]{m}$ is found~\cite{newton07}. The normalization parameter $k$ is then commonly
referred to as $S_\text{1000}$---and given in units of vertical
equivalent muon (VEM)---, and the scaling parameter $r_\text{s}$
is chosen as $r_\text{s} = \unit[700]{m}$. The determination
of the \ac{LDF} is thus reduced to determining $S_\text{1000}$ and
$\beta$.\\

When only data from the ground array are available, the normalization
parameter $S_\text{1000}$ is used to estimate the energy of the
primary particle initiating the recorded air shower. In hybrid events,
however, the primary energy as well as other shower parameters such as
the shower geometry can be determined from the data recorded by the
fluorescence detectors. With this knowledge, we can introduce the
template, or
``photon-optimized'', fit of the \ac{LDF}, which is the basis for the
observable $F_\gamma$. In this fit, we fix the slope parameter $\beta$
to the average value expected for a photon-induced air shower that has the
same primary energy $E$ and the same zenith angle $\theta$ as reconstructed from the
fluorescence detector data. In general, the type of the primary
particle that initiated the recorded air shower is not known. Hence, we use the photon
energy $E_\gamma$, i.e. the calorimetric energy determined
from the fluorescence detector data corrected for the missing energy
expected for a photon-induced air shower (about $\unit[1]{\%}$ of the calorimetric energy~\cite{pierog05}) as reference energy for all events in the
application of the observable.
Since the slope of the \ac{LDF} is fixed in the photon-optimized
\ac{LDF} fit, only the normalization parameter,
which we denote as $S_{1000|\gamma}$, has to be determined from the fit. The function that is
fitted to the data from the ground array thus reads
\begin{equation}
S(r) = S_{1000|\gamma}
\left(\frac{r}{\unit[1000]{m}}\right)^{\beta(E_\gamma, \theta)} \left(\frac{r
    + \unit[700]{m}}{\unit[1700]{m}}\right)^{\beta(E_\gamma, \theta)}.
\end{equation}
Since only a single parameter of the \ac{LDF} has to be determined, the photon-optimized fit can also
be applied to lower-energy events in the EeV range, where perhaps even only one detector from the ground array is
triggered. On a technical note, we use a maximum-likelihood method to
determine the photon-optimized \ac{LDF}, which enables us to use not
only the information from the triggered detectors in the fit, but also
from non-triggering but active detectors. These effectively place an upper
limit on the signal at the corresponding distance to the
shower axis, and it is another advantage of the $F_\gamma$ observable
that this piece of information can be taken into account in a
straightforward way.\\

\begin{figure}[h]
	\centering
		\includegraphics[width=\columnwidth]{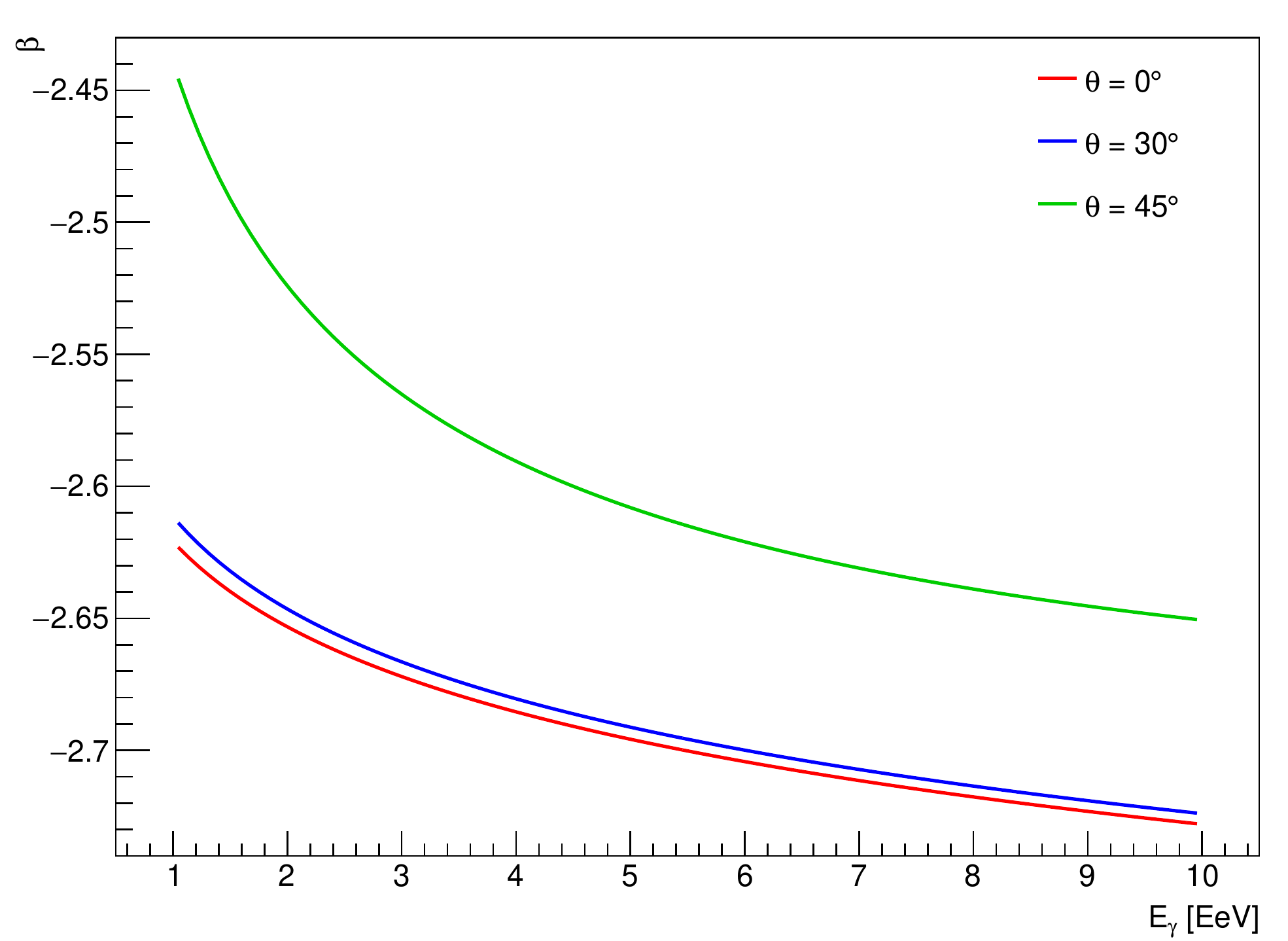}
	\caption{The parameterized slope parameter $\beta$ (following
          Eq.~\ref{eq:betaparameterization}) as a function of the photon
          energy $E_\gamma$ for three different fixed zenith angles $\theta$.}
	\label{fig:betaplot}
\end{figure}

To fix the slope parameter $\beta$, we use a
phenomenological parameterization that has been derived from 
\ac{MC} simulations of photon-induced air showers in the energy range between
$1$ and $\unit[10]{EeV}$ (referring to the true \ac{MC} energy) and in the zenith angle range between $0$ and
$60^\circ$:
\begin{equation}
\beta(E_\gamma, \theta) = a_0(E_\gamma) + a_1(E_\gamma)\,\left(\sec(\theta)-1\right)^3,
\label{eq:betaparameterization}
\end{equation}
with
\begin{equation*}
\begin{split}
a_0(E_\gamma) &= b_0 + b_1\,\log_{10}(E_\gamma\,\unit{[eV]}),\\
a_1(E_\gamma) &= c_0 + c_1\,\log_{10}(E_\gamma\,\unit{[eV]}) + c_2\,\log_{10}(E_\gamma\,\unit{[eV]})^2. 
\end{split} 
\end{equation*}
The five parameters $b_0$, $b_1$, $c_0$, $c_1$ and $c_2$ are:
\begin{equation*}
\begin{split}
b_0 &= -0.695 \pm 0.098,\\
b_1 &= -0.107 \pm 0.005,\\
c_0 &= \phantom{-}506.72 \pm 0.11,\\
c_1 &= -53.159 \pm 0.006,\\
c_2 &= \phantom{-}1.3972 \pm 0.0003.
\end{split} 
\end{equation*}
In Fig.~\ref{fig:betaplot}, the parameterized $\beta$, following
Eq.~\ref{eq:betaparameterization}, is given as a function of the
energy $E_\gamma$ for three different fixed zenith angles $\theta$. Due to
the dependence on secant cubed of the zenith angle, the curves for
$\theta=0^\circ$ and $30^\circ$ are very similar. Only at larger
zenith angles, the curves change significantly.\\

\begin{figure}[h]
	\centering
		\includegraphics[width=\columnwidth]{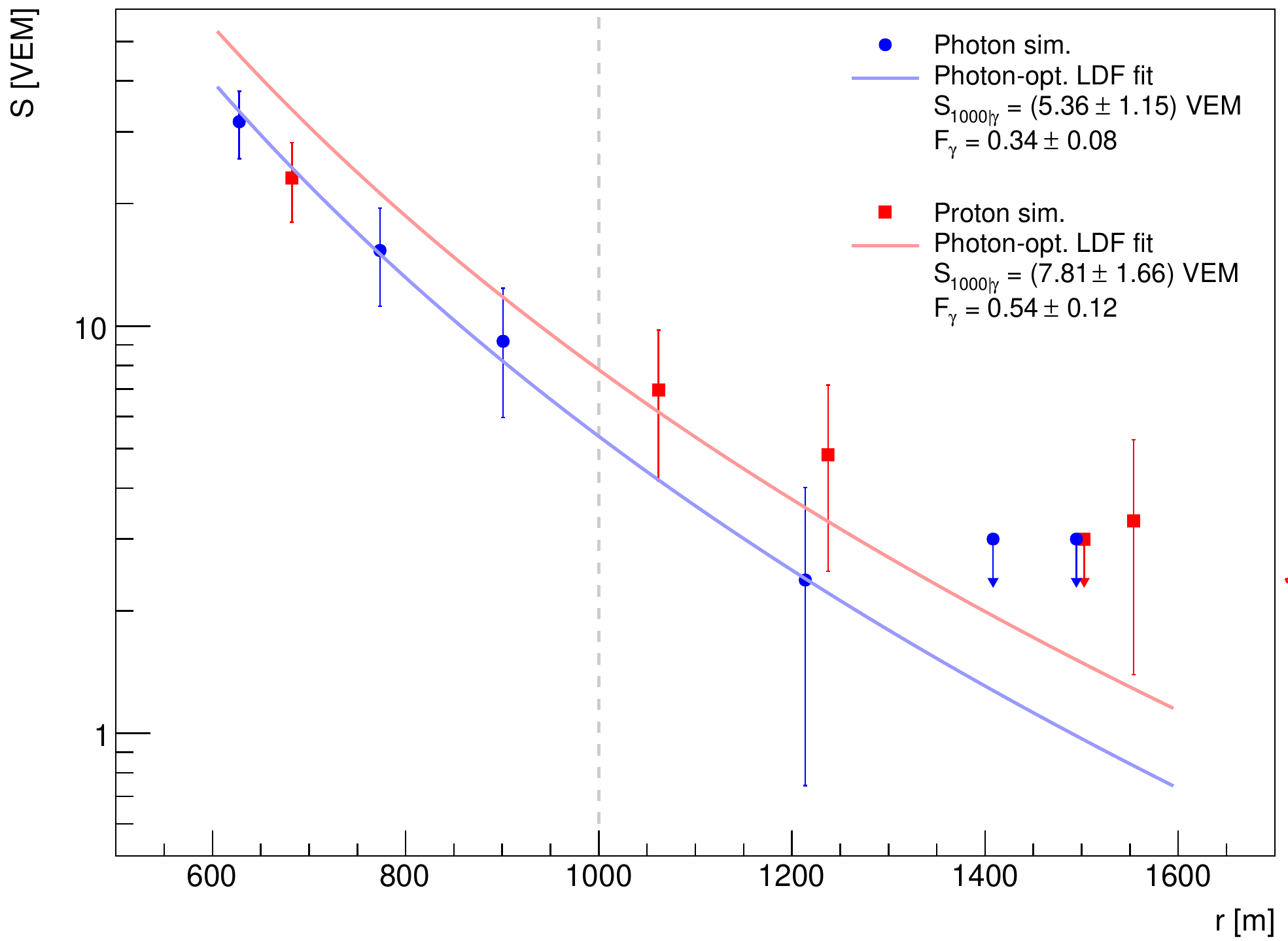}
	\caption{Example for the application of the photon-optimized
          fit of the \ac{LDF} to two simulated air shower events with
          the same primary energy ($E_\text{MC} \sim\unit[3.5]{EeV}$) and zenith
          angle ($\theta \sim 45^\circ$), but different primary
          particle types (photon in blue, proton in
          red). Non-triggering stations that are included in the
          photon-optimized \ac{LDF} fit are indicated as upper limits
          (at $\unit[95]{\%}$ C.L.)
          on the expected signals at the corresponding distances.}
	\label{fig:ldffitexample}
\end{figure}

An example for the application of the photon-optimized
fit to two simulated
air shower events with the same \ac{MC} energy and the same zenith angle---initiated by a primary photon and a primary proton,
respectively---is shown in Fig.~\ref{fig:ldffitexample}. As discussed before, the lateral distribution for the photon-induced
air shower is steeper, leading to a smaller value of
$S_{1000|\gamma}$. It should be pointed out here again that in the
photon-optimized \ac{LDF} template fit, $\beta$ is fixed. Hence, the flatter
\ac{LDF} of a proton-induced air shower is not reproduced
well. However, it is not the purpose of the fit to reproduce the lateral
distribution well and determine the ``true'' $S_{1000}$ for any given
event, but to provide a parameter that helps to discriminate
photon-induced
from hadron-induced air showers.\\

The observable $F_\gamma$ is eventually obtained by normalizing the
$S_{1000|\gamma}$ that is determined from the photon-optimized fit of the \ac{LDF} to the value of $S_{1000}$
that is expected for an average (hadron-induced) air shower with the same primary
energy and the same zenith angle. Thus, $F_\gamma$ is defined as
\begin{equation}
F_\gamma =
\frac{S_{1000|\gamma}}{\left<S_{1000}\right>\!(E_\gamma,\theta)}.
\label{eq:fgammadefinition}
\end{equation}
In hybrid events, the
primary energy $E$ and the zenith angle $\theta$ are determined from the data from
the fluorescence detectors, hence $\left<S_{1000}\right>$
can be calculated from the reconstructed energy and zenith angle by inverting the formulas for the energy
calibration (for the dependence on the energy) and the \ac{CIC}
function (for the dependence on the zenith angle)~\cite{valino15}. As before, we use the photon energy $E_\gamma$ for all
events. For the two example events shown in
Fig.~\ref{fig:ldffitexample}, the $F_\gamma$ values are $0.34\,\pm\,
0.08$ for the photon event and $0.54\,\pm\,0.12$ for the proton event.


\section{Performance}
\label{sec:performance}

\subsection{$F_\gamma$ alone}

To evaluate the performance of the observable for
photon-hadron discrimination, a set of photon- and proton-induced air
showers has been
simulated with CORSIKA~\cite{heck98}, version 7.4000, using
QGSJET-II-04~\cite{ostapchenko11} and
Fluka2011.2b.6~\cite{ferrari05,battistoni07} as hadronic interaction
models at high and low energies, respectively. In total, 20,000 air
showers have been simulated in ten energy bins (equidistant in
$\log_{10}(E_\text{MC})$) in the energy range between 1 and
$\unit[10]{EeV}$ for each of the two primary particle types. As stated
before, we use the setup of the Pierre Auger Observatory as an example for the
application of the observable. Hence, we use the Offline software
framework of the Auger collaboration~\cite{argiro07} to simulate the detector
response of the fluorescence detectors and the individual detectors of
the ground array. For the air shower simulations, we use $10^{-6}$
optimum thinning~\cite{kobal99,risse01}. For the unthinning, we use
the standard routines implemented in the Offline software framework,
which are based on~\cite{billoir08}. Each CORSIKA shower is resampled five times with the
core position of the shower randomly distributed over the area of the
ground array. The total data set of
simulated air shower events contains 200,000 events. Several cuts
are applied to the data set to ensure that only events of
sufficient quality, i.e. with a well-reconstructed geometry and shower
profile following the selection criteria from~\cite{auger17a}, enter the analysis. Overall, this event
selection retains about $\unit[30]{\%}$ ($\unit[43]{\%}$) of all
triggered photon (proton) events at energies around $\unit[1]{EeV}$, increasing to $\unit[42]{\%}$ ($\unit[69]{\%}$)
around $\unit[10]{EeV}$. The differences between the selection
efficiencies for primary photons and protons can be attributed to the
differences between air showers initiated by the two particle types as
discussed in Sec.~\ref{sec:introduction}.\\

\begin{figure*}[p]
	\centering
		\includegraphics[width=0.49\textwidth]{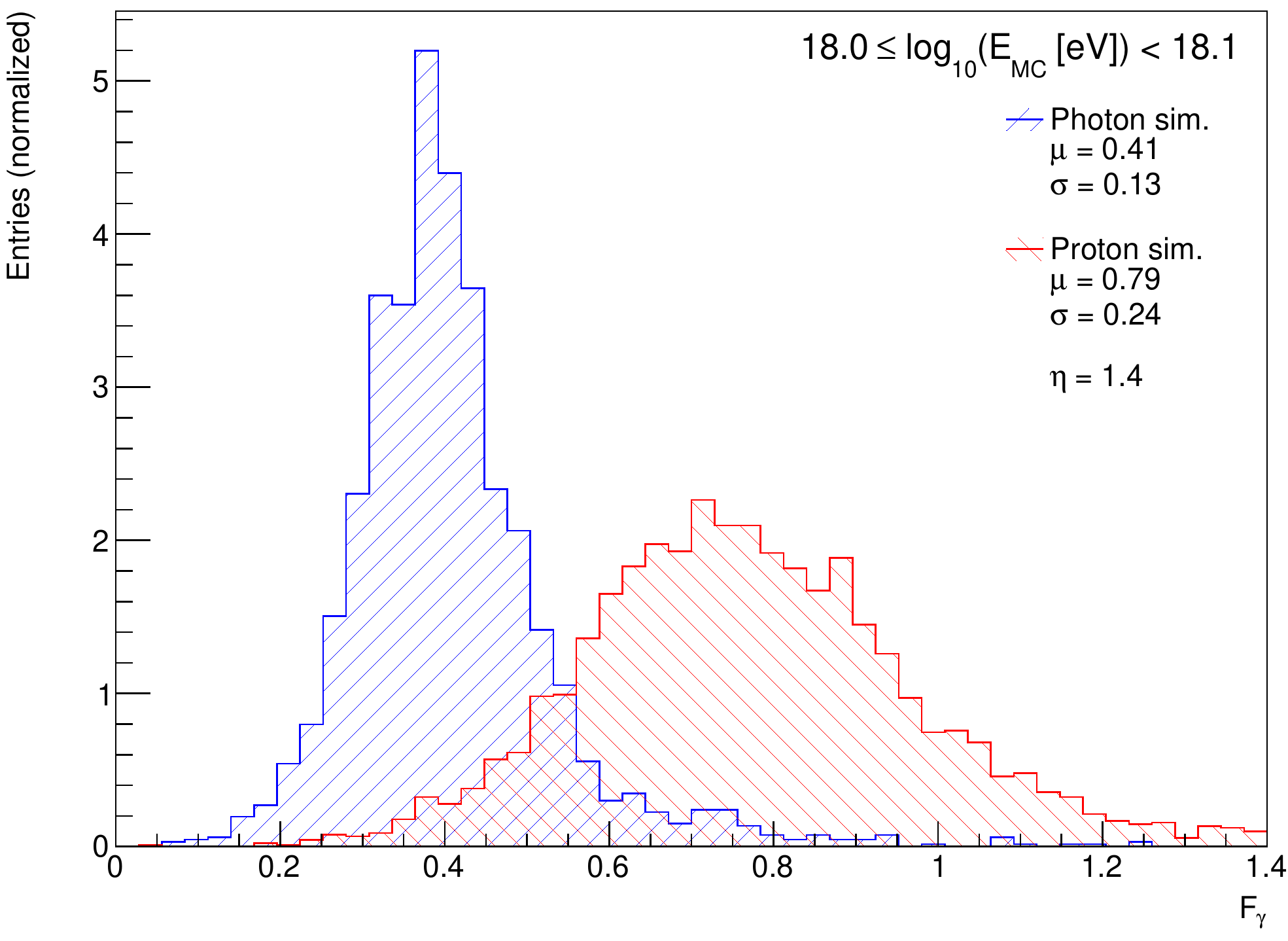}
		\includegraphics[width=0.49\textwidth]{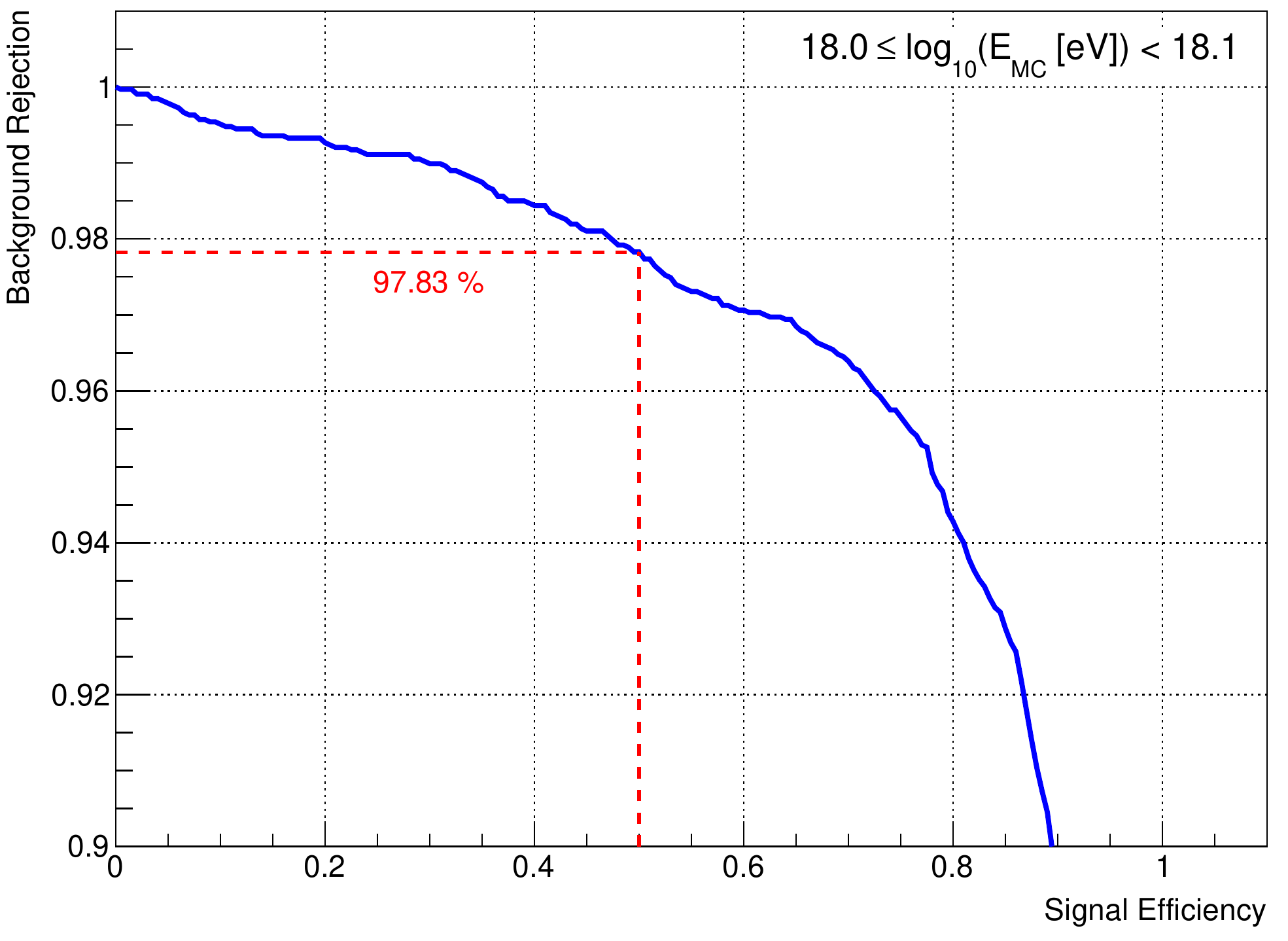}\\
		\includegraphics[width=0.49\textwidth]{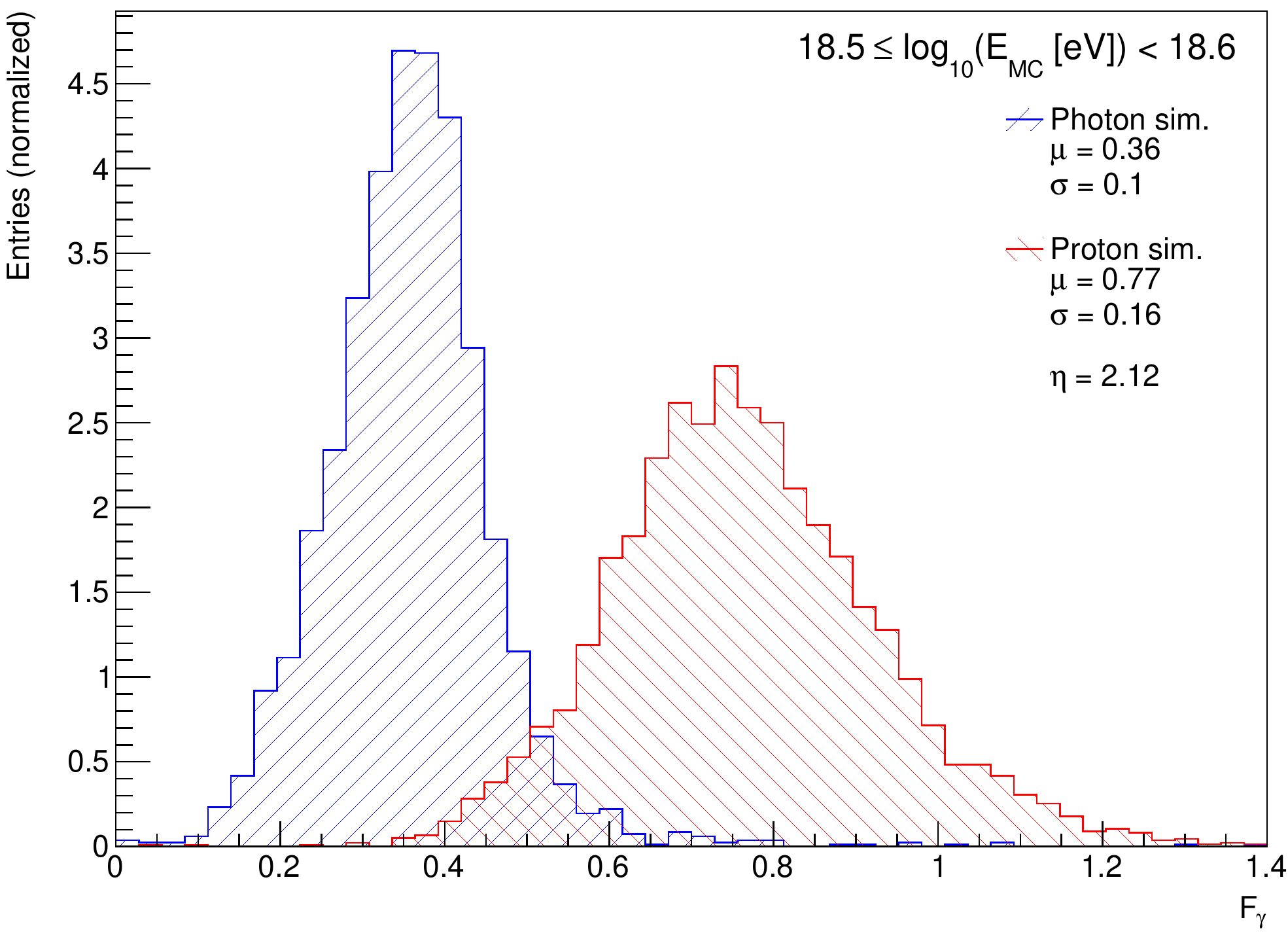}
		\includegraphics[width=0.49\textwidth]{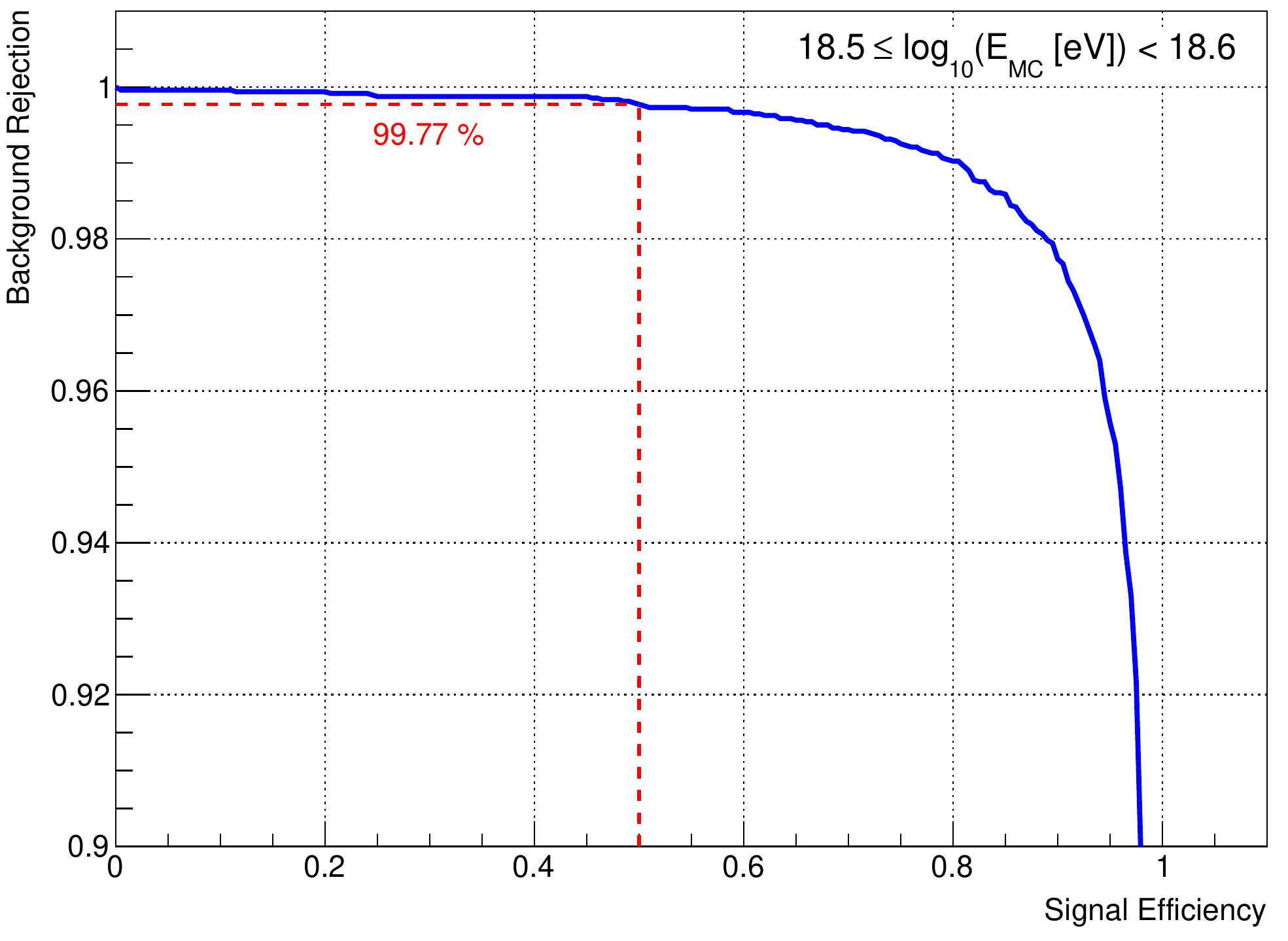}\\
		\includegraphics[width=0.49\textwidth]{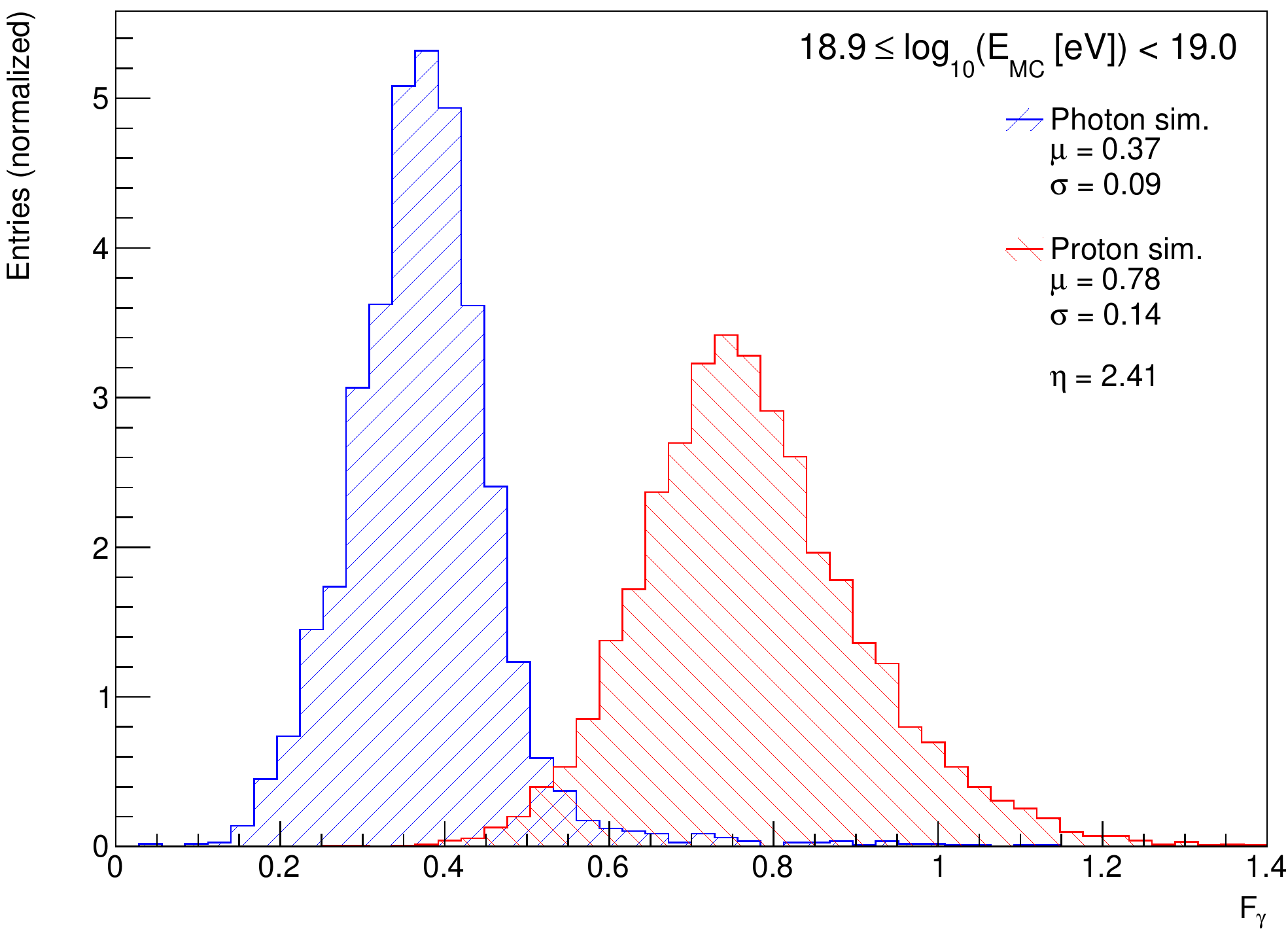}
		\includegraphics[width=0.49\textwidth]{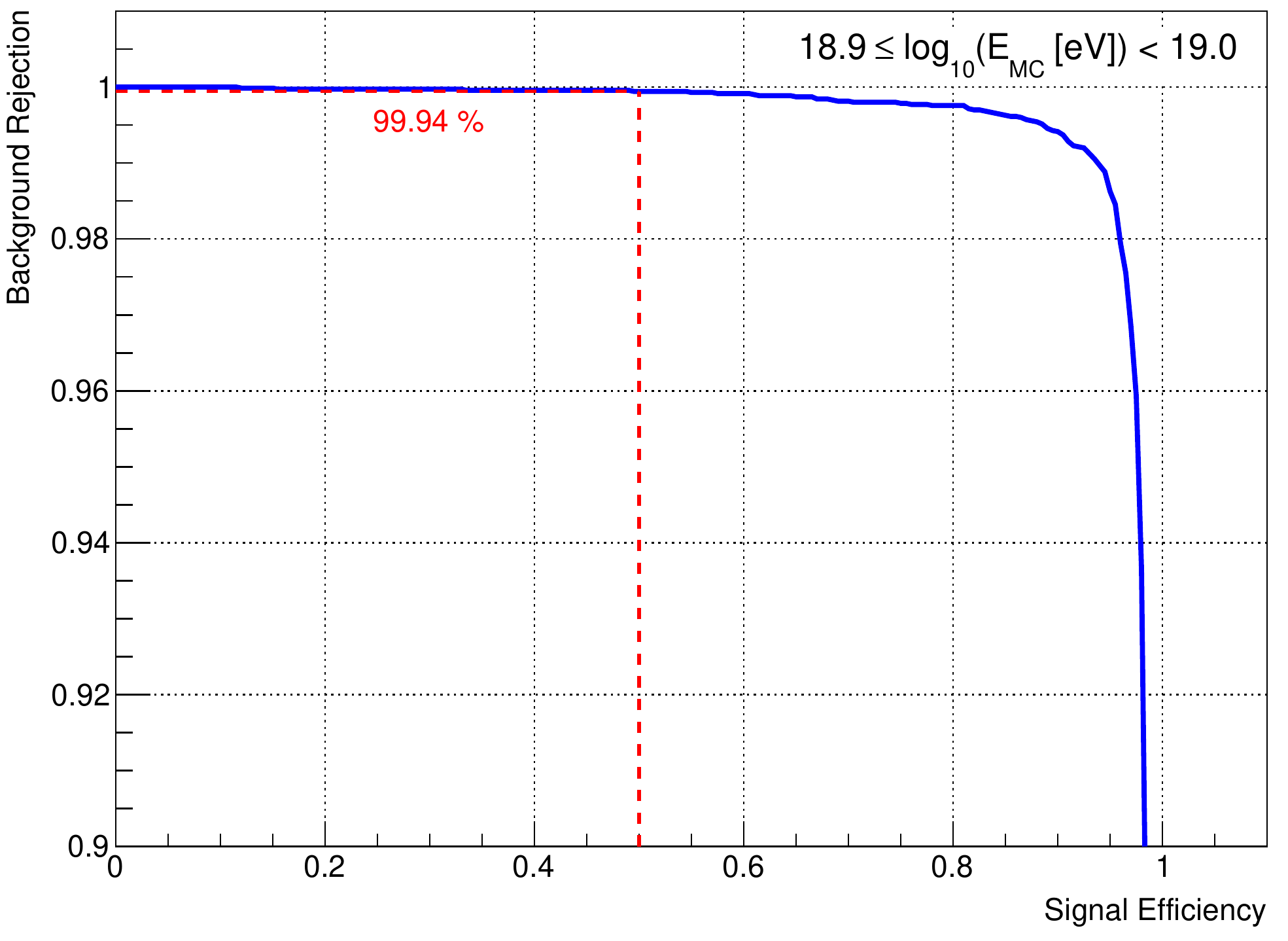}\\
	\caption{Performance of $F_\gamma$ in three different energy
          bins between $E_\text{MC} = \unit[1]{EeV}$ and $E_\text{MC} = \unit[10]{EeV}$. Left column:
          distributions of $F_\gamma$ for primary photons (blue) and
          protons (red). The mean values $\mu$ and the standard deviations
          $\sigma$ of
          the distributions as well as the merit factors $\eta$
          calculated from these values are indicated in the plots. Right column: background rejection
          as a function of the signal efficiency, calculated from the
          $F_\gamma$ distributions. The background rejection at a
          signal efficiency of $\unit[50]{\%}$ is indicated by the
          dashed red lines.}
	\label{fig:fgammaplots}
\end{figure*}

In Fig.~\ref{fig:fgammaplots} (left column), the $F_\gamma$ distributions for three
different energy bins between $E_\text{MC} = \unit[1]{EeV}$ and $E_\text{MC} = \unit[10]{EeV}$ are
shown. The average values of the distributions are
  roughly as expected from the definition of the observable (see
Eq.~\ref{eq:fgammadefinition}): around 0.8 for the proton
distributions (since $S_{1000|\gamma}$ from the photon-optimized
\ac{LDF} fit underestimates the ``true'' $S_{1000}$ by construction, while
$\left<S_{1000}\right>\!(E_\gamma,\theta)$ is close to the ``true''
$S_{1000}$) and around 0.4 for the photon distributions ($S_{1000|\gamma}$ is, on average, close to the ``true'' $S_{1000}$ of photons, but
$\left<S_{1000}\right>\!(E_\gamma,\theta)$ is much larger than the
``true'' $S_{1000}$). Qualitatively, it can be seen that the distributions are well-separated. The separation gets larger with increasing
energy. To quantify the separation between proton- and photon-induced
air shower events, we use the merit factor $\eta$ as a measure for the
separation power of an observable. The merit factor is defined as 
\begin{equation}
\label{eq:eta}
\eta = \frac{\left|\mu_\gamma - \mu_p\right|}{\sqrt{\sigma_\gamma^2 + \sigma_p^2}},
\end{equation}
with $\mu_\gamma$ and $\mu_p$ ($\sigma_\gamma$ and
$\sigma_p$) denoting the mean values (standard deviations) of the photon
and proton distributions, respectively. For the case of the
distributions shown in Fig.~\ref{fig:fgammaplots}, merit factors of
1.40, 2.12 and 2.41 are obtained in the three different energy
bins. The increase in the merit factor comes mostly from the smaller
widths of the distributions at higher energies. With increasing
energy, more stations are triggered and enter the \ac{LDF} fit. Thus,
fluctuations in the signals are mitigated in the fit, leading
eventually to a narrower
$F_\gamma$ distribution. The mean values of the
distributions do not change significantly. This is expected from
the definition of the observable, since the energy dependence is
largely removed by dividing $S_{1000|\gamma}$ by
$\left<S_{1000}\right>\!(E_\gamma,\theta)$.\\

For comparison, the merit factors of the corresponding
$X_\text{max}$ distributions for the same data set have been
calculated. Merit factors of 1.28, 1.76 and 1.97 have been
obtained for the three energy bins. The merit factors calculated here are in good agreement with what is
expected from parameterizations of the $X_\text{max}$ distributions
for primary photons and protons at the corresponding
energies~\cite{dedomenico13}. For the same conditions, the
separation power of $F_\gamma$ in terms of $\eta$ is thus larger than the
separation power of the observable $X_\text{max}$ over the full energy
range considered here. For the observable $S_b$ with $b=4$---the value
that has been used in~\cite{auger17a}---a merit factor around
2 is found in~\cite{ros13} for energies averaged between
$\unit[10^{18.5}]{eV}$  and $\unit[10^{19.6}]{eV}$ and a hexagonal grid with $\unit[1500]{m}$
spacing (cf. the merit factor of 2.41 for $F_\gamma$ around $\unit[10^{19}]{eV}$). For other choices of $b$, the merit factor might
be higher, but it should be noted that in~\cite{ros13}, a
semi-analytical approach was employed instead of a full \ac{MC}
simulation of the detector response.
 Overall, it can be stated that $F_\gamma$ is at
least on par with $S_b$ in terms of the merit factor over the energy
range considered here.\\

It should be noted however, that the merit
factor takes into account only the mean and the width of the
distributions and not the full shape. As an example, one could imagine
a long tail in the
proton distribution reaching beyond the bulk of the photon
distribution. Proton events in this tail will very likely be misidentified
as photon events, even though the bulk of the distributions may be
well-separated, leading to a large merit factor. Hence, we
employ a second measure for the separation power: the background
rejection, i.e. the fraction of events in the proton distribution rejected by a given cut
value on the observable, as a function of the signal efficiency,
i.e. the fraction of events in the photon distribution that pass the
given cut. The resulting curves are shown in
Fig.~\ref{fig:fgammaplots} (right column). As reference value for the
separation power, the
background rejection at a signal efficiency of $\unit[50]{\%}$
(i.e. the cut value corresponds to the median of the photon
distribution) is usually taken. From the three curves shown in
Fig.~\ref{fig:fgammaplots}, values of $\unit[97.83]{\%}$,
$\unit[99.77]{\%}$ and $\unit[99.94]{\%}$ are obtained for the three
energy bins. As before, the corresponding values for the observable
$X_\text{max}$ have also been determined for comparison. In the three
energy bins, values of $\unit[92.60]{\%}$,
$\unit[97.57]{\%}$ and $\unit[98.48]{\%}$ were calculated.\\

\begin{figure}[h]
	\centering
		\includegraphics[width=\columnwidth]{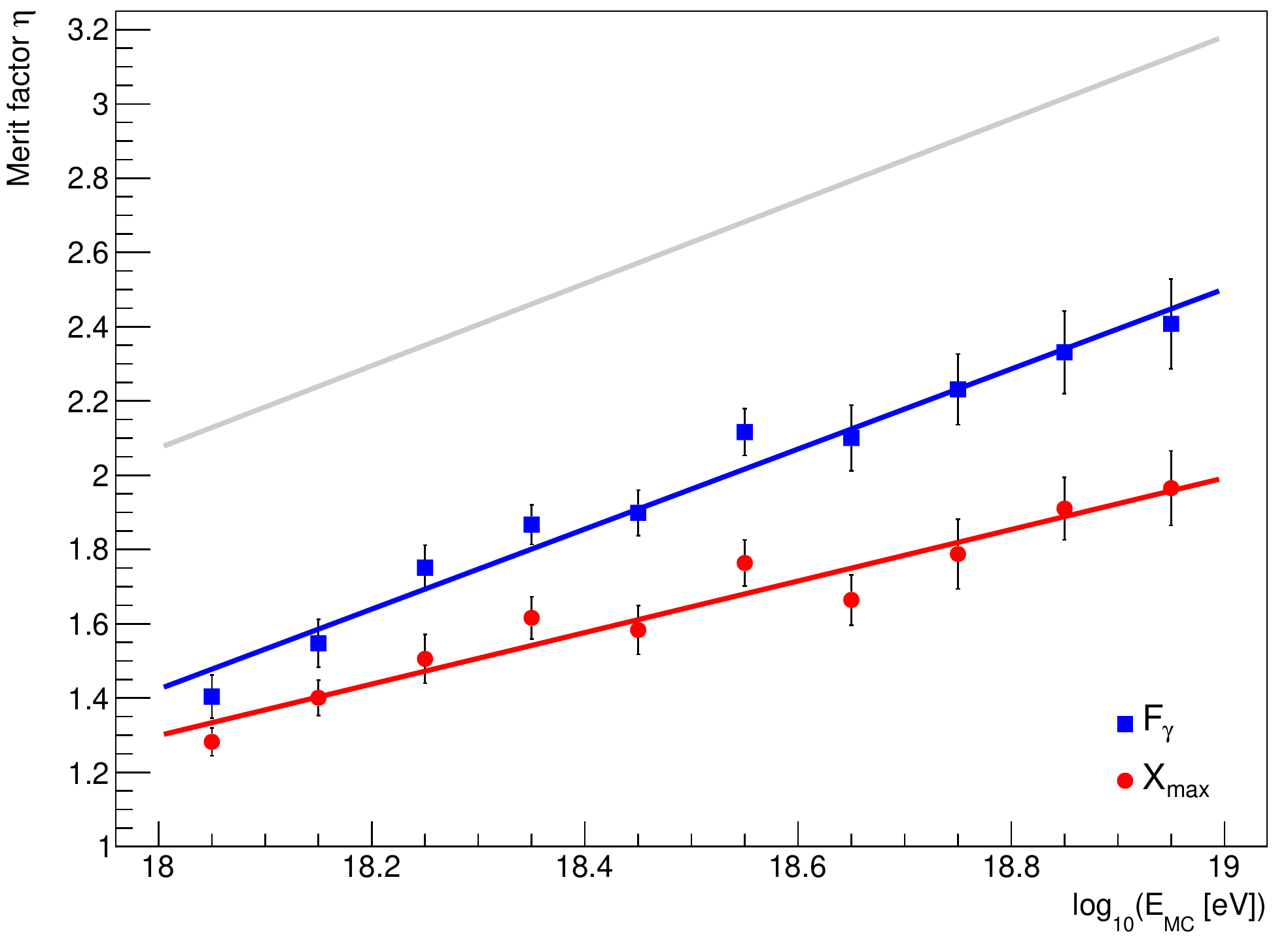}
		\includegraphics[width=\columnwidth]{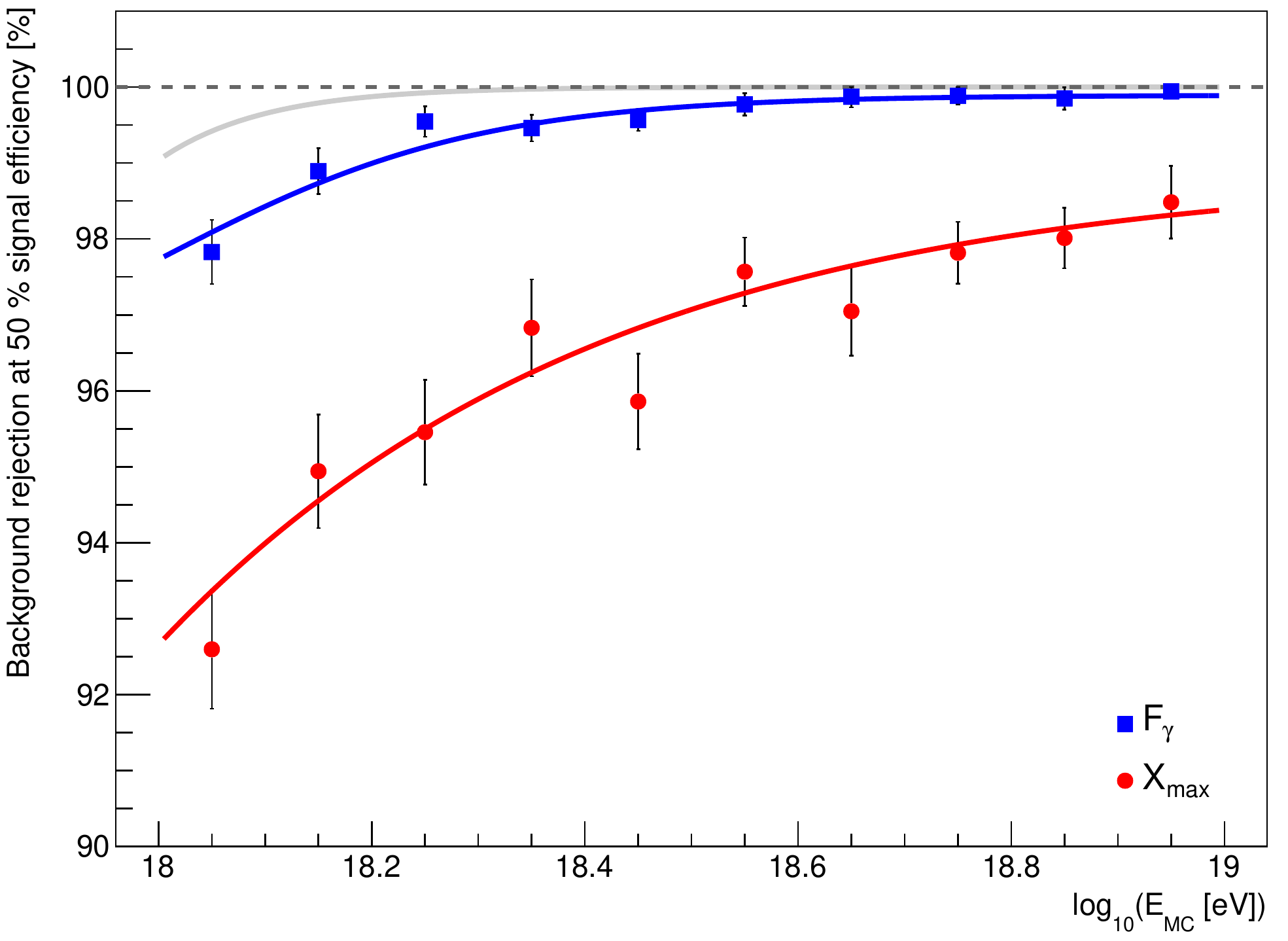}
	\caption{Separation power of $F_\gamma$ (red) between $E_\text{MC} = \unit[1]{EeV}$ and $E_\text{MC} = \unit[10]{EeV}$, compared to the separation power of $X_\text{max}$ (blue). Top: merit factor $\eta$; bottom:
          background rejection $\rho$ at $\unit[50]{\%}$ signal
          efficiency. The uncertainties on $\eta$ and
            $\rho$ have been determined using the bootstrapping
            method~\cite{efron79}. The red and blue lines have been included to
          guide the eye, while the gray lines indicate the separation
          power of a possible combination of the two observables in a
          simple Fisher analysis (cf. Sec.~\ref{sec:combination}).}
	\label{fig:performanceplots}
\end{figure}

In Fig.~\ref{fig:performanceplots}, the performance of the $F_\gamma$
observable in comparison with $X_\text{max}$ is shown over the full
energy range from $1$ to $\unit[10]{EeV}$. In terms of the merit
factor, the separation power of both observables increases linearly
with the logarithm of the energy,
albeit with a larger slope for $F_\gamma$. In terms of the background
rejection, both curves converge exponentially towards
$\unit[100]{\%}$. However, the $F_\gamma$ curve converges much faster:
already around $\unit[10^{18.4}]{eV}$, the curve is above
$\unit[99.5]{\%}$, while the $X_\text{max}$ curve is still below
$\unit[98.5]{\%}$ around $\unit[10^{19}]{eV}$. 

\subsection{Combination with $X_\text{max}$}
\label{sec:combination}

\begin{figure}[h!]
	\centering
		\includegraphics[width=\columnwidth]{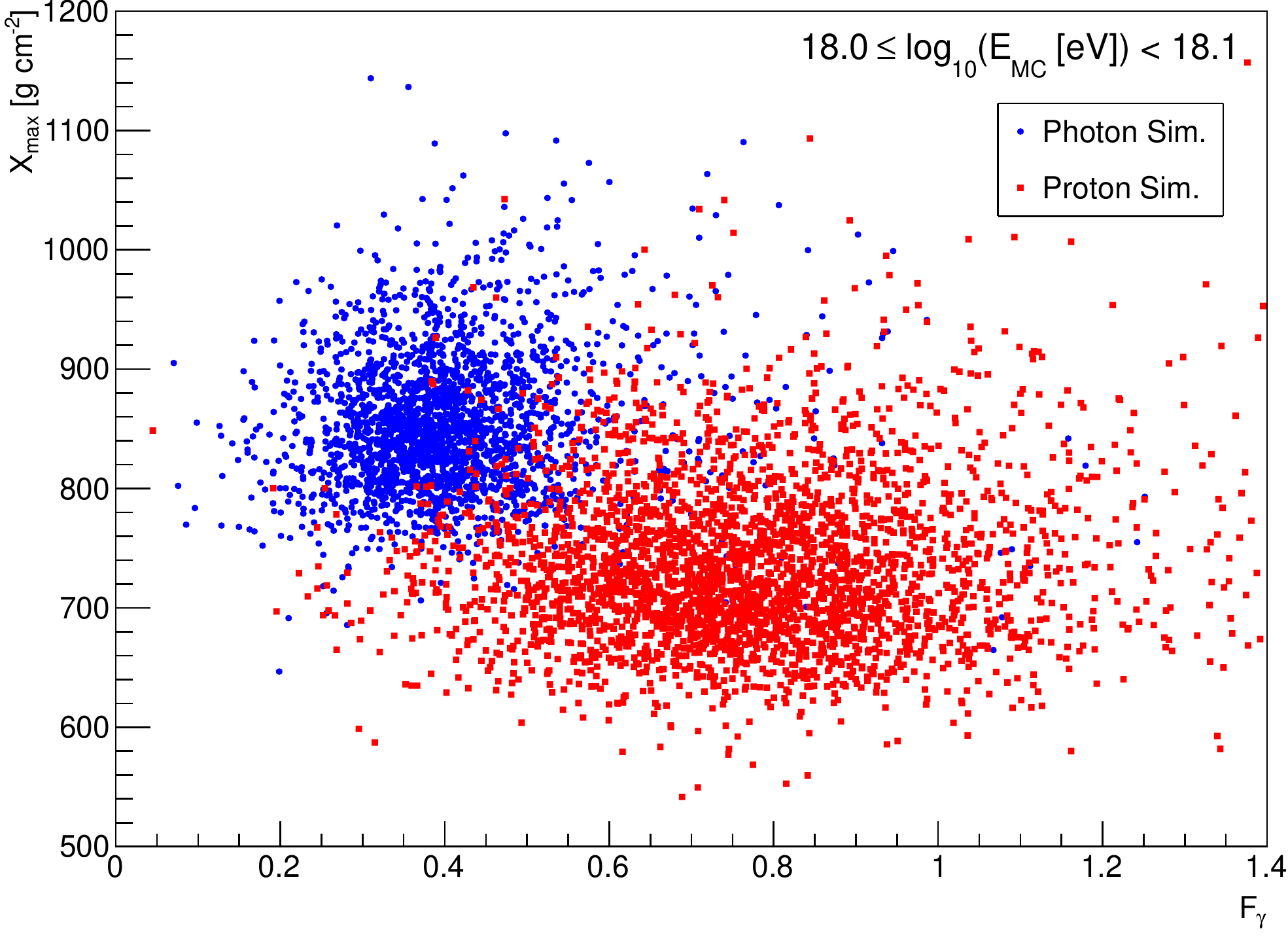}
		\includegraphics[width=\columnwidth]{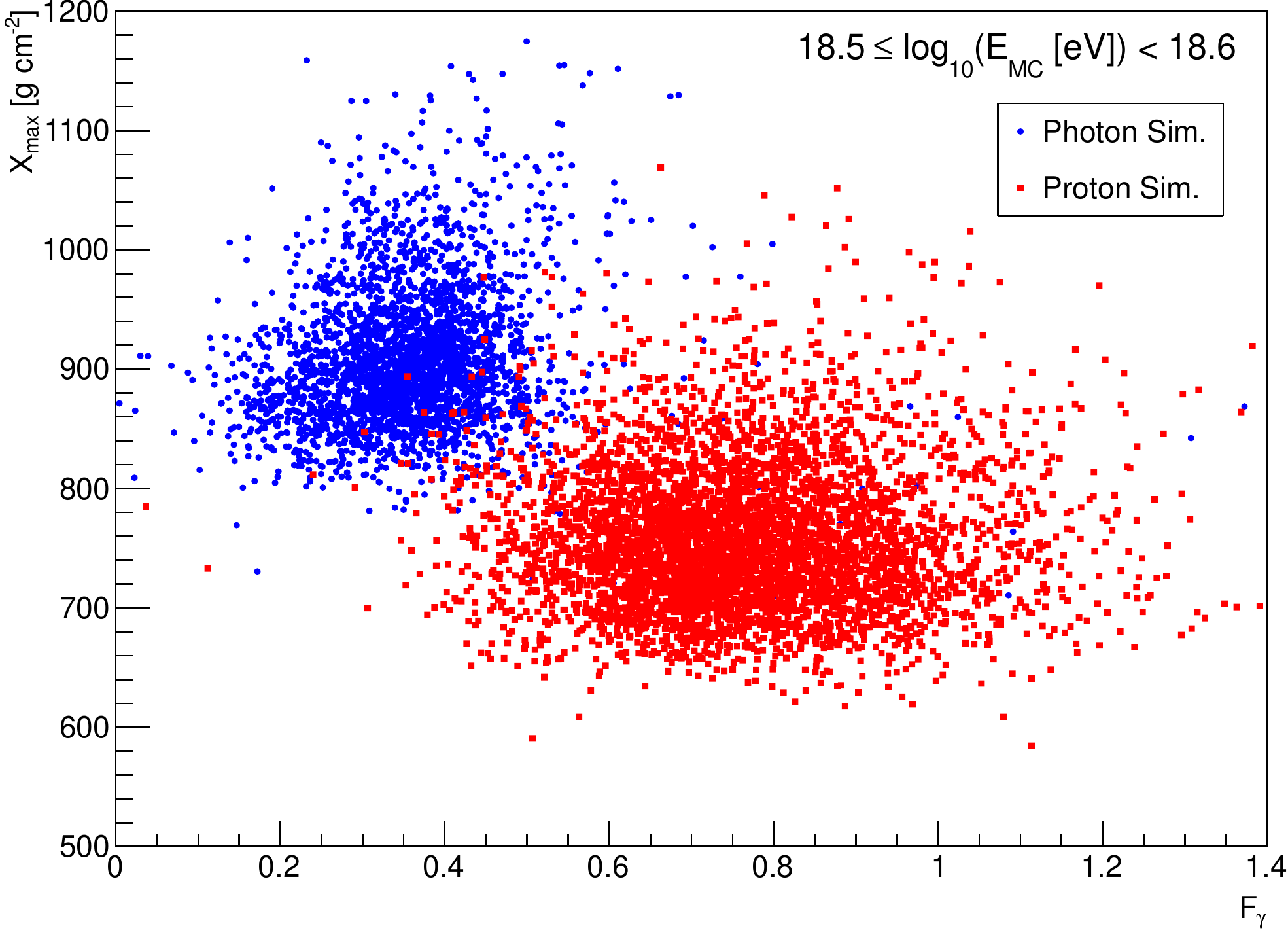}
		\includegraphics[width=\columnwidth]{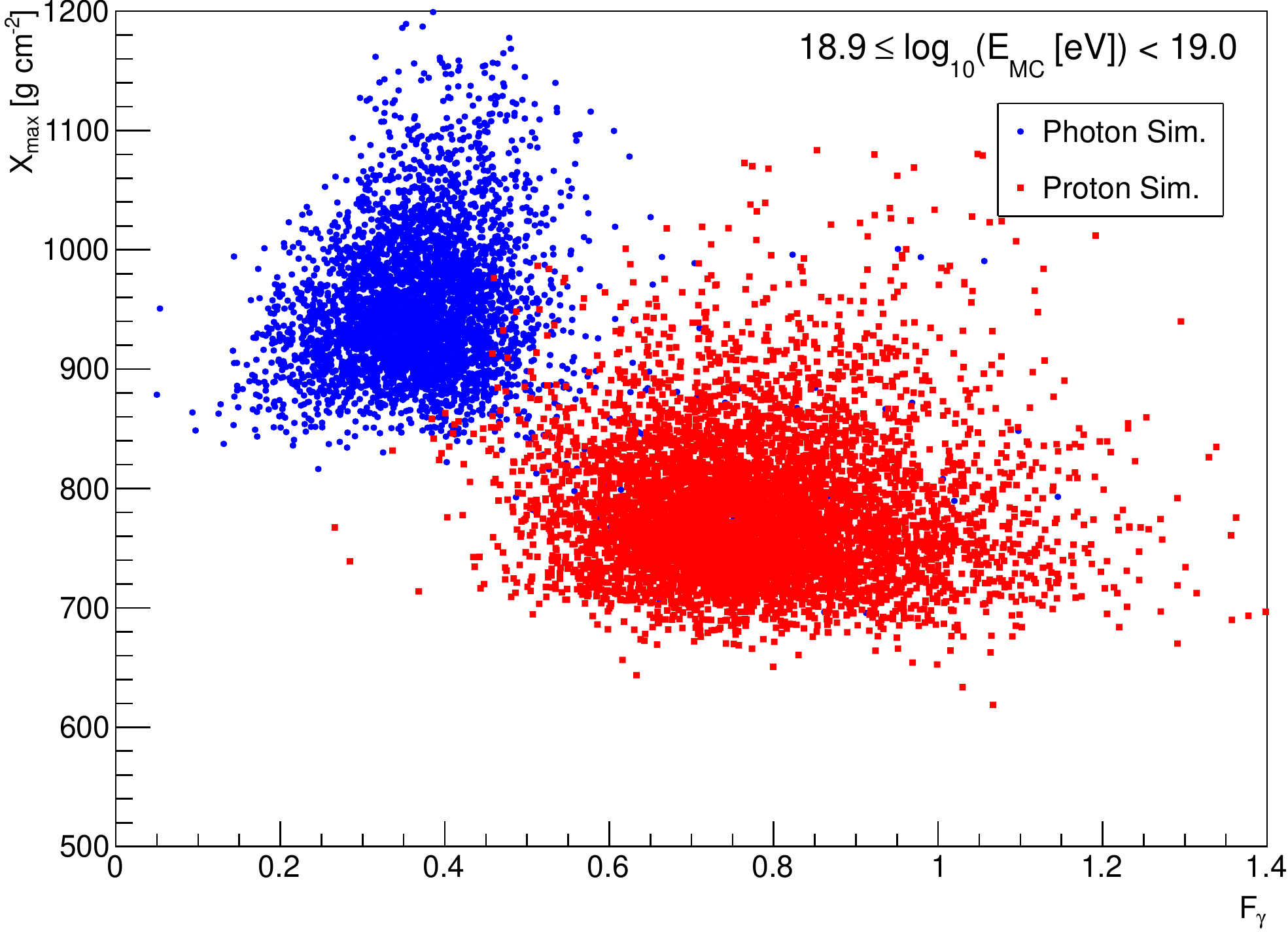}
	\caption{Scatter plots of $F_\gamma$ and $X_\text{max}$ in three different energy
          bins between $E_\text{MC} = \unit[1]{EeV}$ and $E_\text{MC} = \unit[10]{EeV}$ for primary photons (blue) and
          protons (red).}
	\label{fig:scatterplots}
\end{figure}

In a realistic application, $F_\gamma$ will not be used as a
stand-alone observable, but rather in combination with some other
observable, e.g. $X_\text{max}$, to fully exploit the
information available in hybrid events. Hence, we will now discuss the potential performance of the combination of the two
observables $F_\gamma$ and $X_\text{max}$. In
Fig.~\ref{fig:scatterplots}, scatter plots of the two observables in
three different energy bins are shown. Within the individual data sets, the two
observables are largely uncorrelated, with the correlation coefficients
close to zero. Comparing the photon and the proton
data sets, it can be seen that the two observables complement each
other well. The distributions are well separated in the
$(F_\gamma, X_\text{max})$ space, with the separation increasing with
energy as before. Especially at higher energies, only very few proton events are found in the regions
where the bulk of the photon distribution is located. Already by applying
simple cuts on the two observables, it is possible to reach
a background rejection very close to $\unit[100]{\%}$ (i.e. comparable to
the background rejection quoted above for $F_\gamma$), but at a much
larger signal efficiency.\\

\begin{figure}[h]
	\centering
		\includegraphics[width=\columnwidth]{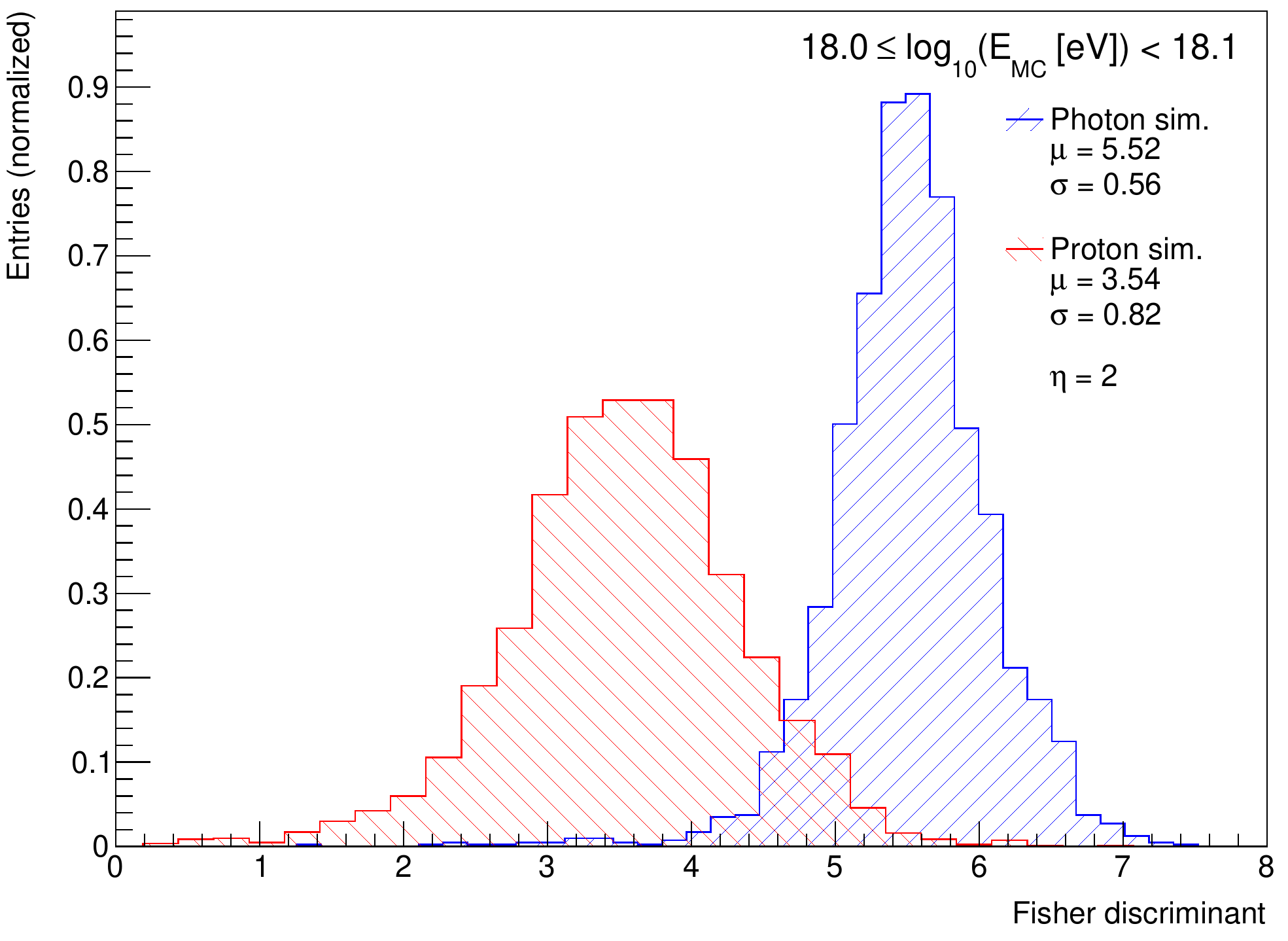}
	\caption{Example for the performance of a combination of $F_\gamma$ and
          $X_\text{max}$ in an \ac{MVA}: distribution of the Fisher discriminant in the first
          energy bin (cf. Fig.~\ref{fig:fgammaplots}).}
	\label{fig:fisherplot}
\end{figure}

Another possibility is to combine the two
observables in a \acf{MVA}. To illustrate the potential of an \ac{MVA}
combining $F_\gamma$ and $X_\text{max}$, we use a simple linear Fisher
discriminant analysis~\cite{fisher36}. The Fisher analysis has the advantages that it
can be calculated analytically and that it provides robust and very
good event classification for uncorrelated input observables, as is the case for $F_\gamma$ and
$X_\text{max}$. In Fig.~\ref{fig:fisherplot}, the distribution of the
Fisher discriminant is shown exemplarily for the first energy bin. As
expected from the scatter plots, the distributions are very well
separated. The overlap between the photon and the proton distributions
is smaller than for $F_\gamma$ alone. Consequently, the merit factor
increases to $2.0$, while the background rejection at a signal
efficiency of $\unit[50]{\%}$ increases to $\unit[99.39]{\%}$. In
other words, the combination of both observables can reduce the
background contamination at this signal efficiency and in this energy
bin by more than a
factor of $3$ compared to $F_\gamma$ alone and more than a factor of
12 compared to $X_\text{max}$ alone. At higher energies, the
combination of the two observables performs similarly well (see also Fig.~\ref{fig:performanceplots}). The merit
factors increase to $2.82$ and $3.10$ in the second and third energy
bin from Fig.~\ref{fig:fgammaplots}, respectively. The values for the background
rejection rise to $\unit[99.94]{\%}$ in the second energy bin and
$\unit[100]{\%}$ (within the limited statistics available in the
data sets used here) in the third energy bin.


\section{Discussion}
\label{sec:discussion}

In summary, we introduced a new observable, called $F_\gamma$, to
improve the photon-hadron separation in air shower events measured
simultaneously by fluorescence detectors and a ground
array. $F_\gamma$ shows very good separation power and complements
nicely to $X_\text{max}$. Combining both observables in an \ac{MVA}
leads to a background rejection of $\unit[99.39]{\%}$ at
$\unit[50]{\%}$ signal efficiency at energies around
$\unit[10^{18}]{eV}$, improving further with increasing energy. In
other words, the observable allows one to reach a nearly background-free
regime while still keeping a significant fraction of the signal. A
particular advantage of $F_\gamma$ is its reconstruction stability: it
can be applied also at energies close to the energy threshold of the
experiment, where perhaps even just a single detector from the ground
array is triggered. The absence of a signal measured in an active
station can be taken into account as well. In addition, the observable is not affected
significantly by
incompletenesses of the ground array, e.g. at the borders of the array
or due to temporarily non-operating detectors. We explicitly checked
this by removing triggered stations from a (simulated) event and repeating
the photon-optimized \ac{LDF} fit without these stations. On average,
the deviation of the resulting
$F_\gamma$ value from the one obtained with all stations included in the
fit is in the order of a few percent only, which is within the average
uncertainties on $F_\gamma$. Especially if only one or two stations
are removed, the deviation is practically negligible. We also note
that the observable is not significantly affected by uncertainties in the hadronic
interaction models used in air shower simulations. This is because the
template used in the photon-optimized \ac{LDF} fit is based only on photon simulations, where the impact of the
choice of the hadronic interaction model is practically negligible
(cf. Fig.~\ref{fig:xmaxplot}). The method of the \ac{LDF} template fit can be easily adapted to other experimental setups than the
case of the Pierre Auger Observatory discussed in this paper. This
includes mixed configurations as well, where e.g. a part of the ground array has
a denser arrangement of detectors.\\

In this work, we discussed the application of the $F_\gamma$ observable for
hybrid data, i.e. data where simultaneous measurements from both a fluorescence
detector and a ground array are available. However, the observable
is not limited to the particular combination of a fluorescence detector and
a ground array. Only the reconstructed shower geometry and primary
energy are needed in addition to the data from the
  ground array to determine $F_\gamma$. This
additional information could also be obtained from e.g. radio
measurements of air showers, which may start to develop into a viable possibility to measure
the longitudinal development of air showers in the atmosphere without 
the limitations in detector uptime that the air-fluorescence technique
faces~\cite{schroeder17}.\\

The method of the \ac{LDF} template fit could also be applied to mass
composition studies, since the characteristic differences between
proton-induced and iron-induced air showers are similar to the
differences between air showers initiated by photons and protons
(i.e. larger $X_\text{max}$, cf. Fig.~\ref{fig:xmaxplot}, and smaller number
of muons). A preliminary study of the proton-iron separation using $F_\gamma$
without any changes to the observable compared to this work leads to
merit factors comparable to those obtained with $X_\text{max}$ at
energies around $\unit[10^{19}]{eV}$. In a real application, the
template used in the \ac{LDF} fit should be adapted, i.e. the fit
should be optimized to e.g. primary iron nuclei. If the appropriate template is
used, it can be expected that the performance of an $F_\gamma$-like
observable for proton-iron separation is even higher.\\

So far, we used a basic event selection to ensure that only
events of sufficient reconstruction quality enter the analysis. The overall photon-hadron separation could be improved further by
introducing additional event selection criteria, optimized for the
specific experimental setup. Such criteria could
be based for example on the goodness of the \ac{LDF} fit. In general,
it is expected that the photon-optimized \ac{LDF} doesn't fit
proton-induced air shower events well (see the two example events
shown in Fig.~\ref{fig:ldffitexample}). Hence, the goodness of the fit
could in principle be used to select photon-like events and to further
improve the search for \ac{UHE} photons. 


\section*{Acknowledgments}
We would like to thank our colleagues from the Pierre Auger Collaboration for many
fruitful discussions. We also thank the anonymous reviewer for his
valuable comments on an earlier version of the manuscript, which
helped to improve the paper. This work is supported by the German Federal Ministry of Education and Research (BMBF) and
the Helmholtz Alliance for Astroparticle Physics (HAP).


\bibliographystyle{elsarticle-num}
\bibliography{references}

\end{document}